\documentstyle[aps,epsf]{revtex}

\begin{document}
\widetext
\title{
Spin-Dependent Coulomb Blockade in
Ferromagnet/Normal-Metal/Ferromagnet Double Tunnel Junctions
}
\author{Hiroshi Imamura}
\address{CREST and Institute for Materials Research, Tohoku University, Sendai
980-8577, Japan}
\author{Saburo Takahashi, and Sadamichi Maekawa }
\address{Institute for Materials Research, Tohoku University, Sendai
980-8577, Japan}
\maketitle
\begin{abstract}
We study theoretically the spin-dependent transport in
ferromagnet/normal-metal/ferromagnet double tunnel junctions by
special attention to cotunneling in the Coulomb blockade
region.  The spin accumulation caused by cotunneling squeezes the 
Coulomb blockade region when the magnetizations in the
ferromagnetic electrodes are antiparallel.
Outside the squeezed Coulomb blockade region, we propose a new
anomalous region, where the sequential tunneling in one of the spin
bands is suppressed by the Coulomb blockade and that in the other is
not.
In this region, the tunnel magnetoresistance oscillates as a function
of bias voltage.
The temperature dependences of the tunnel magnetoresistance and the
magnitude of the spin accumulation are calculated.
\end{abstract} 

\narrowtext
The spin-dependent transport in magnetic materials, in particular, the
tunnel magnetoresistance (TMR) in ferromagnetic junctions,
has attracted much
interest~\cite{julliere75,maekawa82,slonczweski89}.
Recent advances in nano-lithography and thin film processing make it
possible to fabricate very small double tunnel junctions called single
electron transistor (SET),  where the electrostatic energy of excess
electrons in the island has a significant effect on charge
transport, i.e., the Coulomb blockade (CB)~\cite{single_charge_tunneling}. 
In the CB region, 
sequential tunneling, where tunneling events in each
junction occur independently, is
blocked at $T=0$ by the electrostatic energy, and superseded by a more
complex tunneling process called 
``cotunneling''~\cite{averin89,averin90}.
In ferromagnetic SETs \cite{ono97}, 
it was recently pointed out that TMR oscillates in sequential
tunneling regime~\cite{barnas98} and TMR is enhanced in the CB region 
by cotunneling~\cite{takahashi98}.

\begin{figure}
    \epsfxsize=\columnwidth
\centerline{\hbox{
      \epsffile{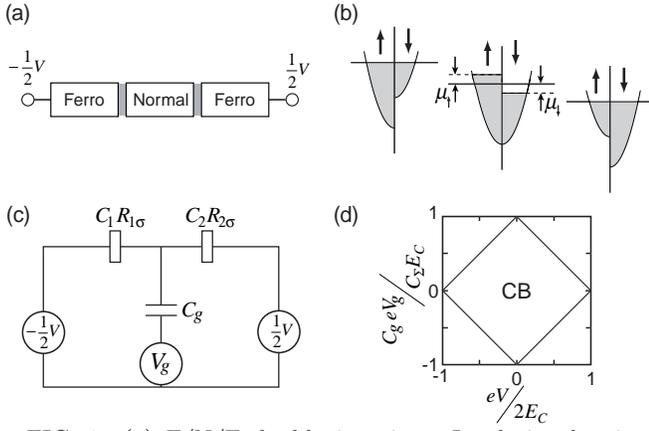}  }}
  \caption{(a) F/N/F double junction. Insulating barriers are shown by the
  shaded rectangles. 
  (b) The densities of states for the
  antiferromagnetic-alignment are schematically shown.
  $\mu_{\uparrow}$ and $\mu_{\downarrow}$ denote the shifts of chemical
  potential for up- and down-spins, respectively.
  (c) F/N/F single electron
  transistor with a gate.
  (d) The Coulomb blockade region for the ferromagnetic-alignment
  with $n=0$.
    }
  \label{fig:cartoons}
\end{figure}

In this Letter, we study theoretically the spin-dependent single
electron tunneling in ferromagnet/normal-metal/ferromagnet (F/N/F)
double tunnel junctions shown in Figs.~\ref{fig:cartoons}(a) and (c),
and find a new anomalous region intrinsic to the spin-accumulated system.
For the antiferromagnetic (A-) alignment, 
where the magnetizations of the right and left electrodes are
anti-parallel as shown in Fig.~\ref{fig:cartoons}(b), electrons
with up(down) spin are easy (difficult) to tunnel into the
normal-metal, and  difficult (easy) to tunnel out of it, because
the densities of states of the left and right electrodes are different
between up and down spin bands. 
This imbalance among the tunnel currents causes the spin
accumulation~\cite{johnson85,valet93,johnson95}, when
the spin-relaxation time is sufficiently
long in the normal-metal.
In the CB region, where sequential tunneling is blocked at $T=0$,
cotunneling causes the spin accumulation.

\begin{figure}
    \epsfxsize=\columnwidth
\centerline{\hbox{
      \epsffile{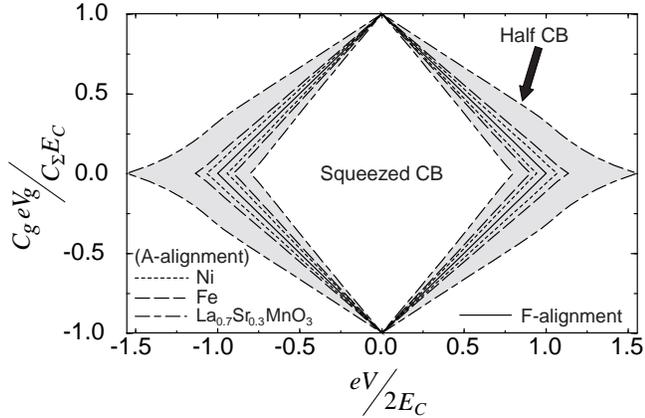}  }}
  \caption{The boundary of the CB region for the
  ferromagnetic-alignment is indicated by the solid line. 
  Boundaries of the half CB region for the
  antiferromagnetic-alignment with Ni, Fe, and
  La$_{0.7}$Sr$_{0.3}$MnO$_{3}$ electrodes are plotted by dotted,
  dashed, and dot-dashed lines, respectively.
  The half CB region for La$_{0.7}$Sr$_{0.3}$MnO$_{3}$
  electrodes is indicated by the shade.
  The tunnel resistance
  for the majority spin band is taken to be  $R_M=2R_{K}$ for
  all systems.
    }
  \label{fig:cb}
\end{figure}

As shown in Fig.~\ref{fig:cartoons}(b), the spin accumulation
increases(decreases) the chemical potential for 
electrons with up(down) spin.
This shift of the chemical potential decreases
the energy for adding(extracting) an electron with down(up) spin
to(from) the normal-metal island.
Therefore the critical voltage, above which
electrons with down(up) spin can tunnel into(out of) the normal-metal
island, is lowered and the CB region is squeezed as shown in
Fig.~\ref{fig:cb}.
On the other hand, the critical voltage above which
electrons with up(down) spin can tunnel into(out of) the normal-metal
island is raised by spin accumulation.
Consequently, a new anomalous region
appears between the lowered and raised critical voltages.
We call this region the half CB region, since the sequential tunneling in
one of the spin bands is suppressed and that in the other is not. 
Cotunneling is essential for appearance of the half CB region.
If cotunneling is not taken into account, 
the spin accumulation does not occur in the CB region and 
the boundary of the CB region does not change, since there is no tunnel
current.
In the half CB region, we find an anomalous oscillation in TMR as a
function of bias-voltage.
The crossover between cotunneling and sequential tunneling~\cite{takahashi98}
is also found in the temperature dependence of the TMR and the spin
accumulation.

The TMR in the ferromagnet/ferromagnet (F/F) single junctions has been
extensively
studied~\cite{julliere75,maekawa82,slonczweski89,miyazaki95,moodera95},
where the tunnel resistance depends on the
relative orientation of the magnetizations between the electrodes,
i.e., parallel or antiparallel.
When the electrodes are made of the same ferromagnetic-metal, the
tunnel resistance for the ferromagnetic (F-) alignment, where the
magnetizations of the left and right electrodes are parallel, is given by 
$1/R_{F}\propto({\cal D}_{M}^{2} + {\cal D}_{m}^{2})$, while for the
A-alignment $1/R_{A}\propto 2{\cal D}_{M}{\cal D}_{m}$,
where ${\cal D}_{M}$ and ${\cal D}_{m}$ are the densities of states for
the majority and minority spin bands at the Fermi level, respectively.
The tunnel magnetoresistance is written as
 \begin{equation}
   \frac{R_{A}-R_{F}}{R_{F}}=\frac{2P^{2}}{1-P^{2}}, 
   \label{eq:tmr_ff}
 \end{equation}
where $P = ({\cal D}_{M}-{\cal D}_{m})/({\cal D}_{M}+{\cal D}_{m})$
is the spin polarization of the electrodes.
The key point is that the difference between the products of the
densities of states for each spin band causes the TMR in F/F junctions.


We now turn to F/N/F double tunnel junctions shown in
Fig.~\ref{fig:cartoons}(a). 
It should be noted that the TMR does not exist in
F/N single junctions,
because there is no difference between the densities of states for up
and down spin bands in the normal-metal.
For F/N/F junctions, however, the TMR exists, if the spin flip
process in the normal-metal island can be neglected.   The origin of
the TMR in F/N/F junctions is quite different  from that for 
F/F junctions.
We assume that the right and left electrodes are made of the same
ferromagnetic-metal such as Ni,  Fe, and  mixed-valent
manganites like La$_{0.7}$Sr$_{0.3}$MnO$_{3}$ and the central
electrode is made of a normal-metal such as Al with sufficiently
long spin relaxation time.

We first consider what happens if we neglect the Coulomb charging
energy.
For the A-alignment, introducing the chemical
potential shift for up(down) spin electrons,
$\mu_{\uparrow}(\mu_{\downarrow})$, 
the spin-resolved tunnel currents through each junction at $T=0$ are
given by
$I_{1\uparrow} \propto {\cal D}_{N} {\cal D}_{M}
(\frac{1}{2}eV-\mu_{\uparrow}),
  I_{1\downarrow} \propto {\cal  D}_{N} {\cal  D}_{m}
(\frac{1}{2}eV-\mu_{\downarrow}),
  I_{2\uparrow} \propto {\cal  D}_{N} {\cal  D}_{m}
(\frac{1}{2}eV+\mu_{\uparrow}),
$ and $
  I_{2\downarrow} \propto {\cal  D}_{N} {\cal  D}_{M}
(\frac{1}{2}eV+\mu_{\downarrow})$.
Here, ${\cal D}_{N}$ is the density of states in the normal-metal for
each spin band,
and subscripts 1 and 2 represent the left and right junctions, 
respectively.

The shifts, $\mu_{\uparrow}$ and
$\mu_{\downarrow}$, are determined by the stability condition, i.e., 
 $I_{1\uparrow} = I_{2\uparrow}$
 and $I_{1\downarrow} = I_{2\downarrow}$.  Then, we have $ \mu_{\uparrow}
= -\mu_{\downarrow} = \frac{1}{2}PeV$.
The tunnel resistance for the A-alignment is written as
$R_{A} \propto (1/{\cal D}_{M} + 1/{\cal D}_{m}) / 4{\cal D}_{N}$. 
For the F-alignment, the spin-resolved currents are
balanced without spin accumulation.
Therefore, the chemical potential for each spin electrons does not shift
and the tunnel resistance is
given by the usual manner as
$R_{F} \propto 1/{\cal D}_{N} ({\cal D}_{M} + {\cal D}_{m})$.
The TMR for F/N/F junctions is obtained as
\begin{equation}
  \frac{R_{A}-R_{F}}{R_{F}}
  =\frac{\left({\cal D}_{M}-{\cal D}_{m}\right)^{2}}{4{\cal D}_{M}
{\cal D}_{m}}
  = \frac{P^{2}}{1-P^{2}},
\label{eq:tmr_film}
\end{equation}
which is half of that for F/F junctions in
Eq.~(\ref{eq:tmr_ff}). 
Equation~(\ref{eq:tmr_film}) has also been obtained by
Brataas {\it et al.}~\cite{brataas98}.
The polarisability $P$ for  Ni, Fe, and
La$_{0.7}$Sr$_{0.3}$MnO$_{3}$ are, respectively, 0.23,
0.40,~\cite{meservey94} and 0.83~\cite{sun97,viret97}.
Introducing these $P$ values into Eq.~(\ref{eq:tmr_film}), we estimate
the TMR for F/N/F double junctions made of Ni, Fe and
La$_{0.7}$Sr$_{0.3}$MnO$_{3}$ electrodes to be 5.6\%,  19\%
and 220\%, respectively.

Let us next examine the single electron transistor with 
a capacitively coupled gate as shown in Fig.~\ref{fig:cartoons}(c).
The transistor is  made of the same materials as the F/N/F double junctions
discussed above, and its size is small enough to observe the
CB.  For simplicity, we assume that the insulating
barriers for junctions 1(left) and 2(right) are the same;
we subsequently set $C_{1}=C_{2}\equiv C$.

The energy changes
due to the forward tunneling of an electron with spin
$\sigma$ through the 
junction 1 $(n\rightarrow n+1)$ and the junction 2 $(n\rightarrow
n-1)$ are given by $E_{1\sigma}^{(+)}(n)
=  E_{C}(1+2n) - \frac{C_{g}}{C_{\Sigma}}eV_{g}
 - \frac{1}{2}eV + \mu_{\sigma}$ and 
$E_{2\sigma}^{(-)}(n) = E_{C}(1-2n) + \frac{C_{g}}{C_{\Sigma}}eV_{g}
 - \frac{1}{2}eV -\mu_{\sigma}$, respectively, 
where $n$ is the number of excess electrons in the normal-metal
island,  $V_{g}$ the gate voltage,
$E_{C} = e^{2}/2C_{\Sigma}$,  and $C_{\Sigma}=2C+C_{g}$.
The energy changes due to the backward tunneling, $E_{1\sigma}^{(-)}(n)$ and
$E_{2\sigma}^{(+)}(n)$, are given by
$E_{1\sigma}^{(-)}(n)=E_{2\sigma}^{(-)}(n)+eV$ and
$E_{2\sigma}^{(+)}(n)=E_{1\sigma}^{(+)}(n)+eV$, respectively.

The total current through the double junctions~\cite{takahashi98} is
expressed as
\begin{equation}
I=\sum_{n=-\infty}^{\infty}
\sum_{\sigma,\sigma^{\prime}=\uparrow\downarrow}
p_{n}\left[I_{1\sigma 2\sigma^{\prime}}(n)
           - I_{2\sigma 1\sigma^{\prime}}(n)
    \right], 
    \label{eq:full_crnt}
\end{equation}
where $p_{n}$ is the probability of charge state $n$ and $I_{j\sigma
k\sigma^{\prime}}(n)$ represents the current, where electrons with spin
$\sigma$ tunnel into the central island through the $j$th junction
and electrons with spin $\sigma^{\prime}$ tunnel out of it through the
$k$th junction.
The probability $p_{n}$ is determined
by the condition for detailed balancing:
$p_{n}[\Gamma_{1}^{(+)}(n) + \Gamma_{2}^{(+)}(n)] =
p_{n+1}[\Gamma_{1}^{(-)}(n+1) + \Gamma_{2}^{(-)}(n+1)]$,
where $\Gamma_{j}^{(\pm)}(n)=\sum_{\sigma}(1/e^{2}R_{j\sigma})
E_{j\sigma}^{(\pm)}(n)/(\exp [E_{j\sigma}^{(\pm)}(n)/T] -1)$ are the
tunneling rates of $n\rightarrow n\pm 1$ in the $j$th junction.
$I_{j \sigma k \sigma ^{\prime}}(n)$ is obtained
by the ``golden rule''~\cite{averin89} as,
\widetext
\begin{eqnarray}
  I_{j \sigma k \sigma ^{\prime}}(n)
 &=& I_{0}
    \frac{1}{R_{j\sigma}R_{k\sigma^{\prime}}}
    \int{\rm d} \epsilon_{1} {\rm d} \epsilon_{2} {\rm d}
    \epsilon_{3}{\rm d} \epsilon_{4} f(\epsilon_{1})
    \left[1- f(\epsilon_{2})\right]
    f(\epsilon_{3}) \left[1- f(\epsilon_{4})\right] \nonumber\\
  &\times&
   \left|
     \frac{1}
      {\epsilon_{2}-\epsilon_{1}+E_{j\sigma}^{(+)}(n)+i\gamma^{(+)}}
    +\frac{1}
      {\epsilon_{3}-\epsilon_{4}+E_{k\sigma^{\prime}}^{(-)}(n)+i\gamma^{(-)}}
   \right|^{2}
   \delta\left(
     \epsilon_{1}-\epsilon_{2}+\epsilon_{3}-\epsilon_{4}+ \Delta E
          \right),
\label{eq:co_current}
\end{eqnarray}
\narrowtext
where $I_{0}\equiv\frac{E_{C}}{e R_{K}}$, $R_{K}={h}/{e^{2}}$ is the
resistance quantum, $R_{j\sigma}$ is the tunnel resistance of the $j$th
junction for electrons with spin $\sigma$, 
$\Delta E$ is the energy difference between initial and final states,
and $ \gamma^{(\pm)}  =\frac{1}{2}\sum_{\sigma}\sum_{j}g_{j\sigma}
  E_{j\sigma}^{(\pm)}(n) \coth [E_{j\sigma}^{(\pm)}(n)/2T]$  
represents the decay rate of the charge state $n\pm1$
with $g_{j\sigma}=\frac{R_{K}}{2\pi R_{j\sigma}}$
~\cite{averin94,averin97}.
Note that in the limit of large tunnel resistance
($g_{j\sigma}\rightarrow 0$) where the cotunneling current is
negligible, Eq.~(\ref{eq:full_crnt}) reduces to the sequential
tunneling current in the orthodox theory~\cite{single_charge_tunneling}.

For the F-alignment, the tunnel resistances
$R_{j\sigma}$ are given by $R_{1\uparrow} = R_{2\uparrow} = R_M$
and $R_{1\downarrow} = R_{2\downarrow} = R_m$, where
$R_M \left(\propto 1/{\cal D}_M{\cal D}_N\right)$ and
$R_m \left(\propto 1/{\cal D}_m{\cal D}_N\right)$ 
are the tunnel resistances for electrons in the majority and minority
spin bands of the ferromagnetic electrodes,
respectively.   The spin does not accumulate, since $I_{i\uparrow
j\downarrow}(n)=$ $I_{i\downarrow j\uparrow}(n)$ for
$\mu_{\uparrow}=\mu_{\downarrow}=0$.
Therefore, the CB region is the same as that for the usual metallic
single electron transistors as shown in Fig.~\ref{fig:cartoons}(d).  

For the A-alignment, the tunnel resistances are given by
$R_{1\uparrow} = R_{2\downarrow} = R_M$ and $R_{1\downarrow} = R_{2\uparrow} = R_m$.
If $\mu_\uparrow=\mu_\downarrow=0$, then
$I_{i\uparrow j\downarrow}(n) \ne I_{i\downarrow j\uparrow}(n)$,
which gives rise to the spin accumulation and non-zero shifts $\mu_{\sigma}$.
The shifts of the chemical potential satisfy the condition that
$\mu_{\uparrow} = - \mu_{\downarrow}$,
because we assumed that the left and right ferromagnetic electrodes
are the same and the density of states in the normal-metal is
constant.
We introduce the symbol
$\delta\equiv\mu_{\uparrow} = -\mu_{\downarrow}$, which is determined by
the stability condition,
 $ \sum_{n}\sum_{i,j}  p_{n}\left[I_{i\uparrow j\downarrow}(n)
     -I_{i\downarrow j\uparrow}(n)\right] = 0 $.
We carry out the numerical integration of Eq.~(\ref{eq:co_current}) and
obtain $\delta$.
Here we take $R_{M}=2R_{K}$ $\left(R_{m}=\frac{1-P}{1+P}R_{M}\right)$.

At $T=0$ and for low $V$ where sequential tunneling is blocked by the CB,
cotunneling causes the spin accumulation in the normal metal island,
leading to the increase (decrease) of the chemical potential of up-spin
(down-spin) electrons by $\delta$ as shown in Fig.~\ref{fig:cartoons}(b).
Therefore, the energy changes due to the single electron
tunneling are given by $E_{i\uparrow}^{(\pm)}(n) =
E_{i}^{(\pm)}(n)\pm\delta$ and  $E_{i\downarrow}^{(\pm)}(n) =
E_{i}^{(\pm)}(n)\mp\delta$, where $E_{i}^{(\pm)}(n)$ are
the energy changes for $\delta=0$. 
The CB region for the A-alignment
is now determined by $E_{1}^{(\pm)}(n)$, $E_{2}^{(\pm)}(n)$
$>\delta$ and squeezed as shown in Fig.~\ref{fig:cb}.
Note that for the F-alignment, the CB region is given by 
$E_{1}^{(\pm)}(n)$, $E_{2}^{(\pm)}(n)$ $>0$.

Outside the squeezed CB region, electrons with 
down(up) spin can tunnel into(out of) the normal-metal
island. However, those with up(down) spin cannot as far as
$E_{1}^{(\pm)}(n)$, $E_{2}^{(\pm)}(n)$ $>-\delta$.
As a consequence, we have a new spin-dependent CB region, where 
the sequential tunneling in one spin bands is suppressed and the other 
is not.
This ``half CB'' region is given by the condition that
$-\delta<$ $E_{1}^{(\pm)}(n)$, $E_{2}^{(\pm)}(n)$ $<\delta$.  
In Fig.~\ref{fig:cb}, the boundaries of this ``half CB'' region for
the A-alignment with  Ni,  Fe and La$_{0.7}$Sr$_{0.3}$MnO$_{3}$
electrodes are plotted for $n=0$.
The solid line indicates the CB boundary for the F-alignment and the
shade represents the half CB region for the A-alignment with
La$_{0.7}$Sr$_{0.3}$MnO$_{3}$ electrodes in Fig.~\ref{fig:cb}.
One can see that the half CB region increases
and  the squeezed CB region, which is surrounded by the
half CB region, decreases as $P$ increases, because $\delta$ increases 
with $P$.

\begin{figure}
    \epsfxsize=\columnwidth
\centerline{\hbox{
      \epsffile{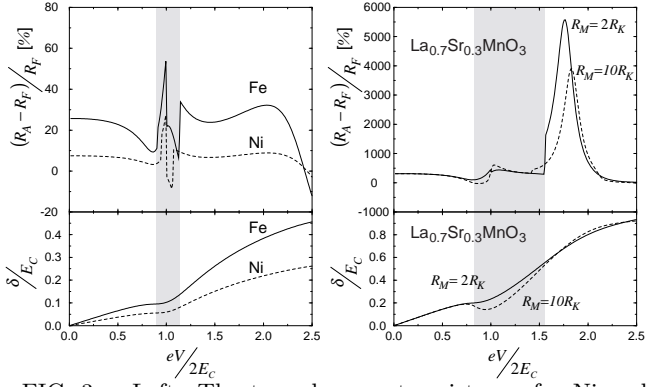}  }}
  \caption{
  Left: The tunnel magnetoresistance for Ni and Fe electrodes are
  plotted against the bias voltage $V$ in the upper panel. 
  The chemical potential shift $\delta$ is
  plotted in the lower panel.  The shaded area represents the
  half CB  region for Fe electrodes.  
  The tunnel resistance for the majority spin electrons
  is $R_M=2R_{K}$, the gate voltage is $V_{g}=0$ and $T=0$.
  Right: The same plot for La$_{0.7}$Sr$_{0.3}$MnO$_{3}$ electrodes.
  The solid(dashed) lines indicate the results for $R_{M}=2R_{K}(10R_{K})$.
  The half Coulomb blockade region for $R_{M}=2R_{K}$ is indicated
  by the shade. 
  The gate voltage is $V_{g}=0$ and $T=0$.
  }
  \label{fig:tmr_delta_zeroT}
\end{figure}

The tunnel magnetoresistance, $(R_{A}-R_{F})/R_{F}$, and the shift
of the chemical potential $\delta$ at $T=0$ and $V_{g}=0$ are plotted in 
Fig.~\ref{fig:tmr_delta_zeroT}.  
At $V\simeq 0$, the TMR for Ni, Fe and
La$_{0.7}$Sr$_{0.3}$MnO$_{3}$ electrodes are about 7.5\%, 26\%
and 310\%, respectively.
The values are $35 \sim 40 \%$ larger than those in Eq.~(\ref{eq:tmr_film}).

As the bias voltage $V$ increases from 0 toward the boundary of the
squeezed CB region, the tunnel current for the
A-alignment increases more
rapidly than that for the F-alignment and the TMR decreases.
At the boundaries of the half CB region, $R_{A}$ jumps
because the current due to cotunneling decreases rapidly and that due
to sequential tunneling starts to flow.
The same jump in $R_{F}$ appears at $eV/2E_{c}=1$, which is the
boundary of the CB region for the F-alignment.
Therefore, the TMR oscillates in the half CB region
as shown in Fig.~\ref{fig:tmr_delta_zeroT}.   
We also find that cotunneling  enhances the TMR around $eV/2E_{c}\simeq 
2.0$, because it suppresses the $V$ dependence of the total current for
the A-alignment.

We have also studied the system with large tunnel resistance,
$R_{M}=10R_{K}$, in order to see what happens when cotunneling is
suppressed.
In Fig.~\ref{fig:tmr_delta_zeroT}, TMR and $\delta$ for
the electrodes of La$_{0.7}$Sr$_{0.3}$MnO$_{3}$ and junctions with
$R_{M}=10R_{K}$ are plotted by the 
dashed lines.
The size of the half CB region is 79\% of that for
$R_{M}=2R_{K}$.

\begin{figure}
    \epsfxsize=\columnwidth
\centerline{\hbox{
      \epsffile{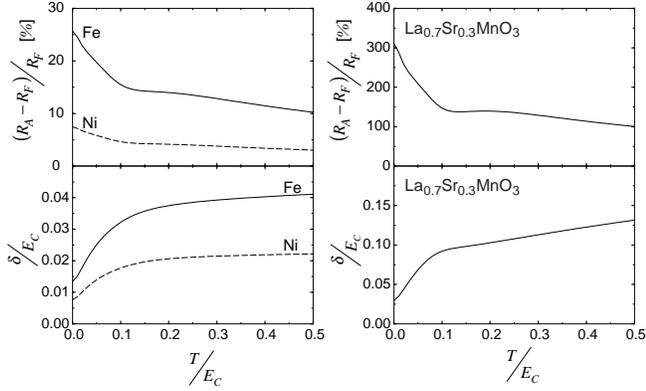}  }}
  \caption{
  Left: The temperature dependence of the tunnel magnetoresistance
  at $V_{g}=0$ and $eV/2E_{c}=0.1$ for Ni and Fe electrodes is shown
  in the upper   panel.  
  The temperature dependence of the chemical   potential shift $\delta$
  at $V_{g}=0$ and $eV/2E_{c}=0.1$ for Ni and Fe electrodes is plotted
  in the lower panel. 
  Right: The same plot for La$_{0.7}$Sr$_{0.3}$MnO$_{3}$ electrodes.
  In both figures the
  tunnel resistance for the majority spin electrons is $R_M=2R_{K}$.
    }
  \label{fig:trm_delta_finiteT}
\end{figure}

The temperature dependences of TMR
and $\delta$ have been calculated at  $V_{g}=0$ and
$eV/2E_{c}=0.1$ in the squeezed CB region.  
As temperature $T$ increases, the sequential tunneling, which is 
exponentially suppressed at low $T$, is recovered.
The results are shown in Fig.~\ref{fig:trm_delta_finiteT}.
One can see clearly the crossover between cotunneling and sequential 
tunneling near $T/E_{c}=0.1$~\cite{takahashi98}.

In conclusion, we have studied the spin-dependent transport in
F/N/F double junctions.  In the antiferromagnetically aligned SET,
cotunneling brings about the spin accumulation at low $V$, 
where sequential tunneling is blocked at $T=0$.
The spin accumulation causes the squeezing of the CB region.  
Outside the squeezed CB region, we have found a new anomalous 
region, where the sequential tunneling in one of the spin bands is
suppressed and that in the other is not.
In this ``half CB'' region, the TMR oscillates as a function of $V$.
The crossover between cotunneling and sequential tunneling
is found in the temperature dependences of TMR and $\delta$.

This work is supported by a Grant-in-Aid for Scientific Research
Priority Area for Ministry of Education, Science and Culture of Japan, 
a Grant for the Japan Society for Promotion of Science, and CREST(Core 
Research for Evolutional Science and Technology Corporation) Japan.

\end{document}